\documentclass[10pt,conference]{IEEEtran}
\IEEEoverridecommandlockouts
\usepackage{graphicx}
\usepackage[table]{xcolor}

\usepackage{cite}
\usepackage{glossaries}
\usepackage[draft, bookmarks=false]{hyperref}
\usepackage[noabbrev,capitalise]{cleveref}
\usepackage{booktabs}
\usepackage{array}
\usepackage[symbol]{footmisc}
\usepackage{threeparttable}
\usepackage{makecell}
\usepackage{arydshln}
\usepackage{tikz}
\usepackage{eso-pic}
\usepackage{hyphenat}
\usepackage{listings}
\usepackage{microtype}
\usepackage{amsmath,amsfonts}
\usepackage{soul}
\usepackage{algorithm}
\usepackage{algpseudocode}
\usepackage{enumitem}
\usepackage{multirow}
\usepackage{xspace}
\usepackage{pifont}
\usepackage{changepage}
\usepackage{multicol}
\usepackage{amssymb}
\usepackage{printlen}
\usepackage{newtxtext,newtxmath}

\makeatletter
\newcommand\fs@spaceruled{%
  \def\@fs@cfont{\bfseries}%
  \let\@fs@capt\floatc@ruled%
  \def\@fs@pre{\vspace{.7\baselineskip}\hrule height.8pt depth0pt \kern2pt}%
  \def\@fs@post{\kern2pt\hrule\relax}%
  \def\@fs@mid{\kern2pt\hrule\kern2pt}%
  \let\@fs@iftopcapt\iftrue}
\makeatother

\usepackage[per-mode=repeated-symbol, free-standing-units, allow-number-unit-breaks]{siunitx}
\DeclareSIUnit\comp{COMP}
\DeclareSIUnit\flop{FLOP}
\DeclareSIUnit\flops{FLOPS}
\DeclareSIUnit\bps{bps}
\DeclareSIUnit\Bps{Bps}
\DeclareSIUnit\gate{GE}
\DeclareSIUnit\op{OP}
\DeclareSIUnit\macu{MACU}
\DeclareSIUnit\ops{OPS}
\DeclareSIUnit\core{core}
\DeclareSIUnit\request{request}
\DeclareSIUnit\cycle{cycle}
\DeclareSIUnit\teraops{TOPS}
\DeclareSIUnit\ghz{GHz}
\DeclareSIUnit\mhz{MHz}
\DeclareSIUnit[number-unit-product = ]\percent{\%}

\definecolor{MidnightBlue}{HTML}{191970}
\definecolor{Mint}{HTML}{3EB889}
\definecolor{EnglishRed}{HTML}{A4515C}
\definecolor{SelectiveYellow}{HTML}{FFBA08}
\definecolor{CyanProcess}{HTML}{08B2E3}
\definecolor{OliveDrab7}{HTML}{4D4730}
\definecolor{Red}{HTML}{FF0000}
\colorlet{color1}{MidnightBlue}
\colorlet{color2}{Mint}
\colorlet{color3}{EnglishRed}
\colorlet{color4}{SelectiveYellow}
\colorlet{color5}{CyanProcess}
\colorlet{color6}{OliveDrab7}
\colorlet{colorAlert}{Red}
\definecolor{PulpGreen}{HTML}{168638}
\definecolor{PulpBlue}{HTML}{1269b0}
\definecolor{PulpRed}{HTML}{a8322c}
\definecolor{PulpYellow}{HTML}{f2c100}

\Crefname{equation}{Eq.}{Eqs.}
\Crefname{figure}{Fig.}{Figs.}
\Crefname{tabular}{Tab.}{Tabs.}

\newacronym{ieee}{IEEE}{IEEE}
\glsunset{ieee}

\makeatletter \newcommand{\AddSpaceIfAnonymous}{\@ifclasswith{acmart}{anonymous}{\vspace{10mm}}{}} \makeatother
\PackageWarning{TODO:}{#1!}

\def\BibTeX{{\rm B\kern-.05em{\sc i\kern-.025em b}\kern-.08em
    T\kern-.1667em\lower.7ex\hbox{E}\kern-.125emX}}

\newacronym{ai}{AI}{Artificial Intelligence}
\newacronym{ml}{ML}{Machine Learning}
\newacronym{nn}{NN}{Neural Network}
\newacronym{cpu}{CPU}{Central Processing Unit}
\newacronym{asic}{ASIC}{Application Specific Integrated Circuit}
\newacronym[longplural={Systems-on-Chip}]{soc}{SoC}{System-on-Chip}
\newacronym{fpga}{FPGA}{Field Programmable Gate Array}
\newacronym{asip}{ASIP}{Application Specific Instruction Processor}
\newacronym{gpp}{GPP}{General Purpose Processor}
\newacronym{gp}{GP}{general-purpose}
\newacronym{gpgpu}{GP-GPU}{General Purpose Graphics Processing Unit}
\newacronym{gpu}{GPU}{Graphics Processing Unit}
\newacronym{sm}{SM}{Streaming Multiprocessor}
\newacronym{cuda}{CUDA}{Compute Unified Device Architecture}
\newacronym{mpi}{MPI}{Message Passing Interface}
\newacronym{cots}{COTS}{Commercial-Off-The-Shelf}
\newacronym{soa}{SoA}{state-of-the-art}
\newacronym{roi}{ROI}{Return on Investments}
\newacronym
[
  longplural={Core Complexes}
]
{cc}{CC}{Core Complex}

\newacronym{lte}{LTE}{Long Term Evolution}
\newacronym{nr}{NR}{New Radio}
\newacronym{4g}{4G}{4th Generation}
\newacronym{5g}{5G}{5th Generation}
\newacronym{b5g}{B5G}{Beyond-5G}
\newacronym{6g}{6G}{6th Generation}
\newacronym{urll}{URLL}{Ultra-Reliable Low-Latency}
\newacronym{mmtc}{mMTC}{massive Machine Type Communications}
\newacronym{embb}{eMBB}{enhanced Mobile Broadband}
\newacronym{3gpp}{3GPP}{3rd Generation Partnership Project}
\newacronym{oran}{O-RAN}{Open-RAN}
\newacronym{ran}{RAN}{Radio Access Networks}
\newacronym{cran}{C-RAN}{Cloud Radio Access Networks}
\newacronym{gnb}{gNB}{Next Generation Node B}
\newacronym{pusch}{PUSCH}{Physical Uplink Shared Channel}
\newacronym{sdr}{SDR}{Software Defined Radio}
\newacronym{phy}{PHY}{Physical}
\newacronym{cu}{CU}{Centralized Unit}
\newacronym{du}{DU}{Distributed Unit}
\newacronym{ru}{RU}{Remote Unit}
\newacronym{ue}{UE}{User Equipment}
\newacronym{ofdm}{OFDM}{Orthogonal Frequency Division Multiplexing}
\newacronym{ofdma}{OFDMA}{Orthogonal Frequency Division Multiple Access}
\newacronym{bf}{BF}{Beam Forming}
\newacronym{mimo}{MIMO}{Multiple-Input, Multiple-Output}
\newacronym{che}{CHE}{Channel Estimation}
\newacronym{dmrs}{DMRS}{Demodulation Reference Signal}
\newacronym{tti}{TTI}{Transmission Time Interval}
\newacronym{sc}{SC}{sub-carrier}
\newacronym{mu}{MU}{Multiple-User}
\newacronym{snr}{SNR}{Signal-to-Noise Ratio}
\newacronym{ber}{BER}{Bit-Error-Rate}

\newacronym{add}{add}{Add}
\newacronym{mul}{mul}{Multiply}
\newacronym{mac}{MAC}{Multiply\&Accumulate}
\newacronym{pmac}{p.mac}{Post-increment Multiply-accumulate}
\newacronym{axpy}{AXPY}{A Times X Plus Y}
\newacronym{dotp}{DOTP}{Dot Product}
\newacronym{sdotp}{SDOTP}{Sum Dot Product}
\newacronym{matmul}{MatMul}{Matrix Multiplication}
\newacronym{gemm}{GEMM}{General Matrix Multiplication}
\newacronym{mvm}{MVM}{Matrix-Vector Multiplication}
\newacronym{fft}{FFT}{Fast Fourier Transform}
\newacronym{sysinv}{SysInv}{Linear System Inversion}
\newacronym{choldec}{CholDec}{Cholesky Decomposition}
\newacronym{mmse}{MMSE}{Minimum Mean Squared Error}
\newacronym{conv2D}{Conv2D}{2D-Convolution}
\newacronym{dct}{DCT}{Direct Cosine Transform}

\newacronym{sram}{SRAM}{Static Random-Access Memory}
\newacronym{dram}{DRAM}{Dynamic Random-Access Memory}
\newacronym{spm}{SPM}{Scratchpad Memory}
\newacronym{tcdm}{TCDM}{Tightly Coupled Data Memory}
\newacronym{IDol}{I\$}{Instruction Cache}
\newacronym{dma}{DMA}{Direct Memory Access}
\newacronym{axi}{AXI}{Advanced eXtensible Interface}
\newacronym{noc}{NoC}{Nework on Chip}
\newacronym{csr}{CSR}{Control Status Register}
\newacronym{hbm}{HBM2E}{High Bandwidth Memory}

\newacronym{ipc}{IPC}{instructions-per-cycle}
\newacronym{wfi}{WFI}{wait-for-interrupt}
\newacronym{raw}{RAW}{read-after-write}
\newacronym{ins}{INS}{instruction}

\newacronym{fpu}{FPU}{Floating Point Unit}
\newacronym{fpss}{FP-SS}{Floating Point Sub-System}
\newacronym{ipu}{IPU}{Integer Processing Unit}
\newacronym{divsqrt}{DIVSQRT}{Division and Square-Root Unit}
\newacronym{lsu}{LSU}{Load Store Unit}
\newacronym{dsp}{DSP}{Digital Signal Processing}
\newacronym{qlr}{QLR}{Queue-Linked Register}

\newacronym{eda}{EDA}{Electronic Design Automation}
\newacronym{ge}{GE}{Gate Equivalent}
\newacronym{fo4}{FO4}{Fan-Out-of-4}
\newacronym{beol}{BEOL}{Back-End-of-Line}
\newacronym{pnr}{PnR}{Place and Route}
\newacronym{ppa}{PPA}{Power, Performance and Area}

\newacronym{numa}{NUMA}{Non-Uniform Memory Access}
\newacronym{fc}{FC}{Fully-Connected}
\newacronym{isa}{ISA}{Instruction Set Architecture}
\newacronym{simd}{SIMD}{Single Instruction Multiple Data}
\newacronym{spmd}{SPMD}{Single Program Multiple Data}
\newacronym{cdf}{CDF}{Cumulative Distribution Function}
\newacronym{api}{API}{Application Programmable Interface}
\newacronym{rtl}{RTL}{Register Transfer Level}
\newacronym{sfr}{SFR}{Synchronization Free Region}
\newacronym{dsl}{DSL}{Domain-Specific Language}

\newacronym{int}{INT}{integer}
\newacronym{fp}{FP}{floating-point}

\newacronym{pe}{PE}{Positional Encoding}
\newacronym{rg}{RG}{Resource Grid}
\newacronym{re}{RE}{Resource Element}
\newacronym{cp}{CP}{Cyclic Prefix}
\newacronym{llr}{LLR}{Log-Likelihood Ratio}
\newacronym{lmmse}{LMMSE}{Linear Minimum Mean Squared Error}
\newacronym{3d}{3D}{3-Dimensional}
\newacronym{2d}{2D}{2-Dimensional}
\newacronym{prb}{PRB}{Physical Resource Block}
\newacronym{mdx}{MDX}{Model Driven Neural Receiver}
\newacronym{dals}{DA-LS}{Data Aided Least Squares}
\newacronym{pals}{PA-LS}{Pilot Aided Least Squares}
\newacronym{cdm}{CDM}{Code-Division Multiplexing}
\newacronym{ldpc}{LDPC}{Low Density Parity Check Code}
\newacronym{bce}{BCE}{Binary Cross-Entropy}
\newacronym{mse}{MSE}{Mean Squared Error}
\newacronym{adam}{ADAM}{Adaptive Moment Estimation}
\newacronym{umi}{UMi}{Urban Microcell}
\newacronym{mcs}{MCS}{Modulation Coding Scheme}
\newacronym{tdl}{TDL}{Tapped Delay Line}
\newacronym{ls}{LS}{Least Squares}
\newacronym{tbler}{TBLER}{Transport Block Error Rate}
\newacronym{cscs}{CSCS}{Swiss National Supercomputing Centre}
\newacronym{flop}{FLOP}{Floating Point Operation}
\newacronym{resblock}{ResBlock}{ResNet block}
\newacronym{qam}{QAM}{Quadrature Amplitude Modulation}

\newacronym{chea}{CHEA}{Channel estimation Attention}

\author{%
Mahdi Abdollahpour\textsuperscript{\ensuremath{\S}}\quad
Marco Bertuletti\textsuperscript{\ensuremath{*}}\quad
Yichao Zhang\textsuperscript{\ensuremath{*}}\quad
Yawei Li\textsuperscript{\ensuremath{\P}}\quad \\
Luca Benini\textsuperscript{\ensuremath{\S*}}\quad
Alessandro Vanelli-Coralli\textsuperscript{\ensuremath{\S*}}
\\
{\small
 \textsuperscript{\ensuremath{\S}}DEI, University of Bologna\quad
 \textsuperscript{\ensuremath{*}}IIS, ETH Z\"{u}rich\quad
 \textsuperscript{\ensuremath{\P}}EEE, Nanyang Technological University 
}
\\
{\small\itshape
\textsuperscript{\ensuremath{\S}}\{mahdi.abdollahpour,luca.benini,alessandro.vanelli\}@unibo.it,
\textsuperscript{\ensuremath{*}}\{mbertuletti,yiczhang,lbenini,avanelli\}@iis.ee.ethz.ch
}
\\[-0.3ex]
{\small\itshape
\textsuperscript{\ensuremath{\P}}li.yawei.ai@gmail.com
}
}

\begin{document}
\title{Scalable Attention for 5G NR Channel Estimation}
\maketitle


\begin{abstract}

Attention-based neural estimators achieve strong channel-estimation accuracy, but the computational cost of global attention over the time-frequency resource grid grows quadratically with the number of subcarriers, and these estimators are typically tied to a single resource allocation.
This paper proposes Channel Estimation Attention (CHEA), a low-complexity channel estimator for 5G New Radio (5G NR) multi-user multiple-input multiple-output (MU-MIMO).
CHEA replaces global attention with a multi-resolution windowed design: a high-resolution encoder preserves local pilot detail, a low-resolution encoder captures wider frequency-domain context, and a local cross-attention decoder transfers this coarse context back to the high-resolution pilot tokens.
A per-Physical Resource Block (PRB) upsampling module then reconstructs the channel over the full slot.
Because every attention operation is confined to a fixed-size window and reconstruction is performed per PRB, the cost of CHEA scales linearly with the number of subcarriers, and a single trained model supports different PRB allocations without retraining.
On a standard-compliant Physical Uplink Shared Channel (PUSCH), CHEA achieves the lowest Mean Squared Error (MSE) among conventional and state-of-the-art neural estimators, while requiring 2.8\(\times\) to 22.0\(\times\) lower operations than existing attention-based estimators.

\end{abstract}



\section{Introduction}
\label{sec:introduction}

\gls{ai}-based physical-layer processing is reshaping how \gls{5g} \gls{nr} receivers acquire channel state information.
Learned channel estimators now routinely surpass the classical \gls{ls} and \gls{lmmse} baselines that have long served \gls{ofdm} systems~\cite{honkala2021deeprx,cammerer2023neural,wiesmayr2024design,mdx2025}.
Sustaining these gains in a deployed uplink receiver, however, is constrained less by accuracy than by two practical requirements: the estimator must fit the tight latency, memory, and compute budget of the \gls{gnb}, and it must operate across the many different \gls{pusch} resource allocations that users are scheduled with~\cite{mdx2025}.

In \gls{pusch}, the receiver estimates the \gls{mimo} channel from \glspl{dmrs} placed sparsely over the time-frequency resource grid, and the quality of this estimate directly affects \gls{mimo} detection, residual interference after equalization, and link-level reliability.
Classical estimators remain attractive because of their simplicity and interpretability.
The \gls{ls} estimator divides the received pilots by the known \glspl{dmrs} and interpolates over the remaining resource elements, but it ignores channel correlation and is sensitive to noise.
The \gls{lmmse} estimator improves accuracy using second-order channel statistics and the noise variance, yet it requires reliable covariance information and matrix operations whose cost grows with the cube of the resource-grid size~\cite{savaux2017lmmse}.
These limitations motivate learning-based estimators that infer the full channel response by exploiting the time-frequency structure of the observed pilots~\cite{gizzini2022survey}.

Among data-driven estimators, convolutional networks exploit local time-frequency patterns to denoise and interpolate the pilot-domain channel, while attention-based models capture the non-local correlations that purely local operations cannot represent.
Transformer-style estimators such as HA02~\cite{luan2022attention}, Channelformer~\cite{luan2023channelformer}, and the vision-transformer-based CEViT~\cite{liu2024pd} report accuracy well beyond classical interpolation, and lightweight convolutional designs such as InterpolateNet~\cite{luan2021low} pursue the same goal at lower cost.

Two obstacles limit these estimators in practice.
First, global attention has a computational cost that grows quadratically with the number of subcarriers; for wideband \gls{pusch} allocations this quickly becomes prohibitive on a resource-constrained receiver.
Second, most neural estimators are trained for a single, fixed bandwidth, so a separate model must be retrained and stored for every \gls{prb} allocation, which is at odds with the flexibility a scheduler actually requires.
An estimator that is both scalable in bandwidth and reusable across allocations is therefore still missing.

In this paper, we propose \gls{chea}, a scalable attention-based channel estimator for \gls{mimo} \gls{pusch} reception that addresses both problems.
Instead of attending over the full resource grid, \gls{chea} processes the pilot-domain channel in fixed-size frequency windows: a high-resolution encoder preserves fine local pilot structure, a low-resolution encoder captures wider frequency-domain context, and a local cross-attention decoder transfers the coarse context back to the high-resolution pilot tokens.
A per-\gls{prb} upsampling layer then reconstructs the full-slot channel.
Because every attention operation is confined to a fixed-size window and the reconstruction is performed per \gls{prb}, the complexity of \gls{chea} scales linearly with the number of subcarriers, and a single trained model supports a wide range of \gls{prb} allocations without bandwidth-specific retraining.
Simulation results on a standard-compliant \gls{5g} \gls{nr} \gls{pusch} chain show that \gls{chea} attains the lowest \gls{mse} among the evaluated classical, convolutional, and \gls{soa} attention-based estimators, while its \gls{mac} count scales linearly with the number of subcarriers and its parameter count stays fixed at about $36$K across all allocations. 
For the 264-subcarrier allocation, this corresponds to ${\approx}1.7$M \glspl{mac}, which is $2.8\times$ to $22.0\times$ fewer \glspl{mac} and $5.9\times$ to $38.6\times$ fewer parameters than the attention-based methods. Our implementation will be available in our repository at~\cite{open_source}.
The main contributions of this work are summarized as:

\begin{itemize}
    \item We propose \gls{chea}, a scalable attention-based channel estimator for \gls{5g} \gls{nr} \gls{mimo} \gls{pusch} whose complexity scales linearly with the number of subcarriers.
    \item We design a multi-resolution windowed attention architecture, coupled with a local high-/low-resolution cross-attention decoder, that captures both local pilot detail and wider frequency-domain context without global attention.
    \item We benchmark \gls{chea} against \gls{ls}, \gls{lmmse}, InterpolateNet, HA02, Channelformer, and CEViT in terms of both \gls{mse} and computational complexity.
\end{itemize}

\vspace{-7pt} 

\section{System Model}
\label{sec:system_model}
\vspace{-2pt} 

We consider a \gls{5g} \gls{nr} MU-\gls{mimo} \gls{pusch} uplink in which $N_{TX}$ single-layer \glspl{ue} are co-scheduled on the same time-frequency resources and jointly received by a \gls{gnb} with $N_R$ receive antennas.
As each \gls{ue} sends a single layer, the index $n_{TX}\in\{1,\dots,N_{TX}\}$ identifies both a layer and its \gls{ue}.
Each slot occupies a time-frequency resource grid $\mathcal{RG} = \{1,\dots,F\}\times\{1,\dots,S\}$,
where $F$ is the number of subcarriers and $S$ the number of \gls{ofdm} symbols.
A resource element is indexed by $(f,s)\in\mathcal{RG}$, with $f$ the subcarrier index and $s$ the \gls{ofdm}-symbol index.
The frequency axis is partitioned into \glspl{prb} of $N_{\mathrm{SC}}^{\mathrm{PRB}}=12$ subcarriers each, so that $F = 12\,N_{\mathrm{PRB}}$ for $N_{\mathrm{PRB}}$ allocated \glspl{prb}.

Let $\mathbf{x}_{f,s}\in\mathbb{C}^{N_{TX}}$ stack the symbols transmitted by the $N_{TX}$ \glspl{ue} at resource element $(f,s)$, and let $\mathbf{y}_{f,s}\in\mathbb{C}^{N_R}$ be the received vector after FFT processing.
The received vector superimposes the contributions of all co-scheduled \glspl{ue},
\vspace{-5pt} 
\begin{equation}
    \mathbf{y}_{f,s}
    =
    \mathbf{H}_{f,s}\,\mathbf{x}_{f,s}
    +
    \mathbf{n}_{f,s},
    \label{eq:rx_signal_model}
\vspace{-4pt} 
\end{equation}
where the $n_{TX}$-th column of $\mathbf{H}_{f,s}\in\mathbb{C}^{N_R\times N_{TX}}$ is the channel from \gls{ue} $n_{TX}$ to the $N_R$ receive antennas, and $\mathbf{n}_{f,s}\sim\mathcal{CN}(\mathbf{0},N_0\mathbf{I}_{N_R})$ is additive complex Gaussian noise with variance $N_0$.

Channel estimation relies on \glspl{dmrs} placed sparsely over the grid.
The \gls{dmrs} resource elements of \gls{ue} $n_{TX}$ form the set $
    \mathcal{P}_{n_{TX}}
    =
    \{(f,s)\in\mathcal{RG} : (f,s)\text{ carries DMRS for UE }n_{TX}\},
$
and the complete pilot set is $\mathcal{P}=\bigcup_{n_{TX}=1}^{N_{TX}}\mathcal{P}_{n_{TX}}$.
In the considered \gls{pusch} configuration, \glspl{dmrs} occupy two \gls{ofdm} symbols given by the pilot-symbol set $\mathcal{S}_p=\{2,11\}$, and \glspl{ue} sharing the same pilot resource elements are separated by a frequency-domain \gls{cdm} group of size $2$.
The estimation task is to recover the full-grid channel $\{\mathbf{H}_{f,s}\}_{(f,s)\in\mathcal{D}}$ of all \glspl{ue} from the pilot observations on $\mathcal{P}$, where $\mathcal{D}\subset\mathcal{RG}$ denotes the data-carrying resource elements.

For learning-based processing, the channel of each link $\ell=(n_R,n_{TX})$ is represented by stacking its real and imaginary parts along the innermost dimension
\begin{equation}
    \mathbf{H}_{\ell}^{\mathbb{R}} \in \mathbb{R}^{F\times S\times 2}.
    \label{eq:real_channel_tensor}
\end{equation}
Estimation accuracy is reported as the \gls{mse} over all links and data-carrying resource elements,
\vspace{-4pt} 
\begin{equation}
    {\mathrm{MSE}}
    =
    \frac{1}{|\mathcal{D}|\,N_R N_{TX}}
    \sum_{(f,s)\in\mathcal{D}}
    \bigl\|
    \hat{\mathbf{H}}_{f,s}-\mathbf{H}_{f,s}
    \bigr\|_F^2 .
    \label{eq:mse_loss}
    \vspace{-6pt} 
\end{equation}


\subsection{Baseline Channel Estimators}
\label{sec:baseline_estimators}
\vspace{-4pt} 

We compare \gls{chea} against conventional estimators and against \gls{soa} neural channel estimators.

\subsubsection{\gls{ls} Estimation}
At the pilot locations, the \gls{ls} estimator computes the channel coefficient by dividing the received pilot observation by the known transmitted \gls{dmrs} symbol. 
The \gls{ls} estimate for link $(n_R,n_{TX})$ is
\vspace{-5pt} 
\begin{equation}
    \hat{h}^{\mathrm{LS}}_{f,s,n_R,n_{TX}}
    =
    \frac{
    p^{*}_{f,s,n_{TX}}y_{f,s,n_R}
    }{
    |p_{f,s,n_{TX}}|^2
    },
    \qquad (f,s)\in\mathcal{P}_{n_{TX}},
    \label{eq:ls_estimator}
\vspace{-5pt} 
\end{equation}
where $p_{f,s,n_{TX}}$ is the known \gls{dmrs} symbol. 
When \gls{cdm} groups are used, the pilot estimates belonging to the same orthogonal group are averaged. Then the estimates~\eqref{eq:ls_estimator} can be interpolated to the full grid as $\hat{\mathbf{H}}^{\mathrm{LS}}_{\ell}\in\mathbb{C}^{F\times S}$.

\subsubsection{\gls{lmmse} Estimation}
The \gls{lmmse} estimator improves the pilot-domain estimate by exploiting channel correlation and noise variance. 
Let $\hat{\mathbf{h}}^{\mathrm{LS}}_{\mathcal{P}}$ denote the vectorized \gls{ls} pilot estimate for a given antenna-layer link, and let $\mathbf{h}$ denote the vectorized full-grid channel. 
The \gls{lmmse} estimate is
\vspace{-5pt} 
\begin{equation}
    \hat{\mathbf{h}}^{\mathrm{LMMSE}}
    =
    \mathbf{R}_{\mathbf{h}\mathbf{h}_{\mathcal{P}}}
    \left(
    \mathbf{R}_{\mathbf{h}_{\mathcal{P}}\mathbf{h}_{\mathcal{P}}}
    +
    \mathbf{R}_{\mathbf{e}}
    \right)^{-1}
    \hat{\mathbf{h}}^{\mathrm{LS}}_{\mathcal{P}},
    \label{eq:lmmse_estimator}
\vspace{-5pt} 
\end{equation}
where $\mathbf{R}_{\mathbf{h}\mathbf{h}_{\mathcal{P}}}$ is the cross-covariance between the full channel and pilot-domain channel, $\mathbf{R}_{\mathbf{h}_{\mathcal{P}}\mathbf{h}_{\mathcal{P}}}$ is the pilot-domain channel covariance, and $\mathbf{R}_{\mathbf{e}}$ is the covariance of the \gls{ls} estimation error. 
In Sionna~\cite{hoydis2022sionna}, the \gls{lmmse} interpolation baseline applies covariance-aware filtering over the resource grid, using the configured frequency, time, and spatial covariance matrices~\cite{hoydis2022sionna}. 
Although this estimator is a strong model-based baseline, its complexity and memory footprint increase with cube of the grid size, and its performance depends on the accuracy of the assumed covariance model. Sionna implementation uses $10^6$ \gls{umi} channel samples to estimate the covariance matrices. 


\subsubsection{Neural Methods}
Beyond conventional model-based estimators, we compare \gls{chea} with several
learning-based estimators, including \gls{soa} attention-based methods:
\begin{itemize}
    \item \textit{InterpolateNet}: a lightweight convolutional residual network that refines interpolated pilot-domain estimates. It has low parameter count and relies mostly on local convolutional structure~\cite{luan2021low}.
    \item \textit{HA02}: it uses transformer-style processing to improve the use of pilot observations~\cite{luan2022attention}.
    \item \textit{CEViT}: vision-transformer-based channel estimators that use patch embedding and attention to capture time-frequency channel correlations~\cite{liu2024pd}. To have a fair comparison of the methods, the extra \gls{snr}, Doppler, and delay spread tokens are ignored.
    \item \textit{Channelformer}: an encoder-decoder architecture using multi-head attention as an input pre-processor followed by a residual convolutional decoder~\cite{luan2023channelformer}. 
\end{itemize}

\vspace{-2pt} 

These baselines provide a comparison between purely model-based estimators, convolutional neural estimators, and global attention-based estimators.

\vspace{-8pt} 

\section{Channel Estimation Attention (CHEA)}
\label{sec:chea}
\vspace{-4pt} 

\begin{figure}[htbp]
\vspace{-10pt} 
\centerline{\includegraphics[width=.77\columnwidth]{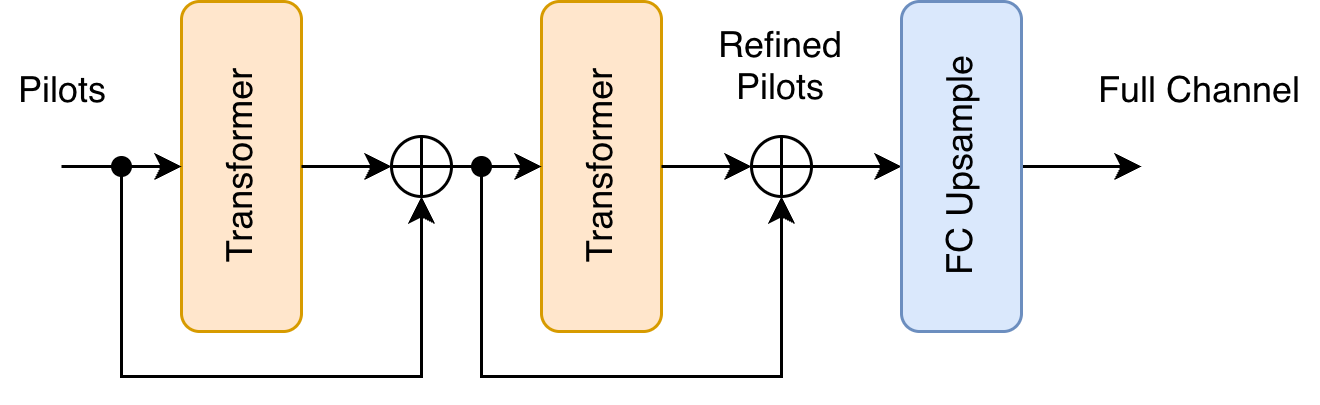}}
\vspace{-11pt}
\caption{Channel estimation block diagram.}
\label{fig:framework_nn}
\vspace{-15pt}
\end{figure}

\begin{figure}[t]
\centerline{\includegraphics[width=.99\columnwidth]{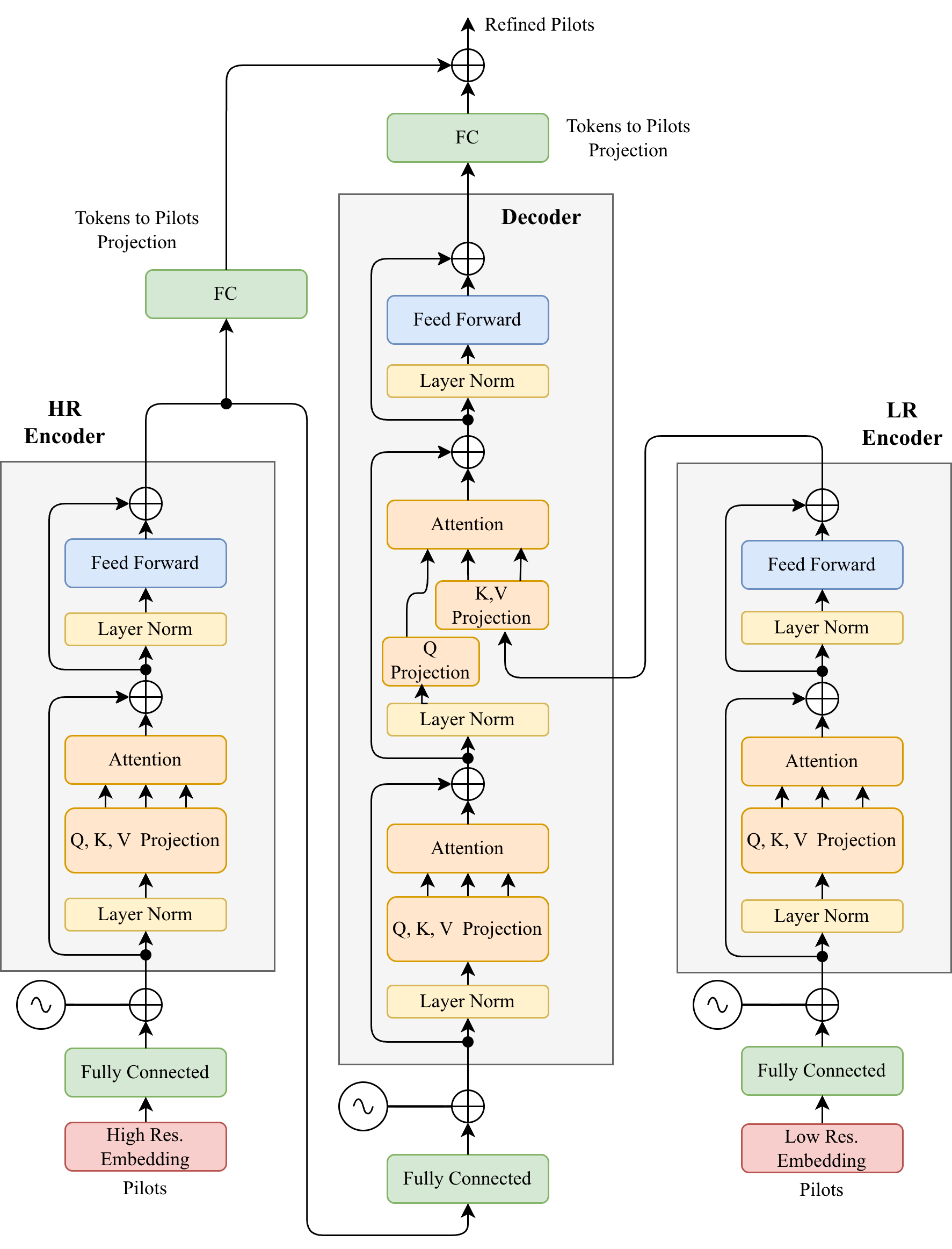}}
\vspace{-13pt}
\caption{CHEA transformer block diagram.}
\label{fig:framework_transformer}
\vspace{-15pt}
\end{figure}

This section introduces \gls{chea}, the proposed Channel Estimation Attention transformer.
As illustrated in Fig.~\ref{fig:framework_nn}, the model consists of a high-resolution encoder, a low-resolution encoder, and a cross-attention decoder, followed by a per-\gls{prb} upsampling block.
It estimates the full \gls{mimo} channel over a \gls{pusch} resource grid from pilot-domain observations while maintaining linear complexity in the number of subcarriers.
The two encoders and the decoder use the same transformer building blocks but \emph{do not share weights}; each module has its own independent parameters.
The architecture of a transformer block is shown in Fig.~\ref{fig:framework_transformer}.

\subsection{Input Representation}
\label{subsec:chea_input}
\vspace{-3pt}

\gls{chea} is applied independently to each antenna-layer link, so the link dimension is merged with the batch dimension during training and inference.
For each link, the input is the \gls{ls} estimate interpolated only in the frequency axis with its real and imaginary parts stacked along the innermost dimension,
\vspace{-5pt}
\begin{equation}
    \mathbf{X}_{p}\in\mathbb{R}^{F\times|\mathcal{S}_p|\times2}.
    \label{eq:input_tensor}
\vspace{-5pt}
\end{equation}

\subsection{Multi-Resolution Windowing}
\label{subsec:multi_resolution_windowing}

\gls{chea} processes the pilot tensor with two parallel branches that use the same tokenization but operate at different frequency resolutions:
\begin{itemize}
    \item a \emph{high-resolution} branch that preserves local pilot detail over short windows;
    \item a \emph{low-resolution} branch that captures wider frequency context after average pooling.
\end{itemize}
The two branches are later combined by the cross-attention decoder.
We next describe the common tokenization, the low-resolution pooling, and the window partitioning.

\paragraph{Tokenization}
Tokenization is a reshape only; no convolution or dense layer is applied at this stage.
Each \gls{prb} spans $P=12$ subcarriers and carries, for a given link, $|\mathcal{S}_p|$ pilot \gls{ofdm} symbols with real and imaginary parts.
Each \gls{prb} is split into $2|\mathcal{S}_p|$ tokens, one per (pilot-symbol, real/imaginary) pair, and each token is the length-$P$ vector collecting the $P=12$ subcarrier values of that pair.
Stacking the $F/P$ \glspl{prb} and merging the antenna-layer links into the batch dimension gives the tokenized tensor
\vspace{-7pt}
\begin{equation}
    \mathbf{X}_{\mathrm{tok}}
    \in
    \mathbb{R}^{B \times \frac{F}{P} \times 2|\mathcal{S}_p| \times P},
\end{equation}
where $B$ is the batch size after link merging.

\paragraph{Low-resolution pooling}
The low-resolution branch averages every $\rho=4$ neighboring \gls{prb} tokens along the frequency axis, so that each low-resolution patch summarizes $\rho$ adjacent \glspl{prb} while remaining a length-$P$ vector, similar to the average-pooled anchors in~\cite{li2023efficient}.
This reduces the number of frequency patches from $F/P$ to $F/(\rho P)$, retaining wideband context at a lower token density.

\paragraph{Window partitioning}
Each branch groups its patches into fixed-size frequency windows.
The high-resolution branch uses windows of $N_h=6$ \glspl{prb}, while the low-resolution branch uses windows of $N_l=24$ \glspl{prb}, with subcarrier lengths
\vspace{-4pt} 
\begin{equation}
    W_h=N_hP=72,
    \qquad
    W_l=N_lP=288.
\end{equation}
The number of windows in each branch is
\begin{equation}
    N_{w,h}=\left\lceil\frac{F}{W_h}\right\rceil,
    \qquad
    N_{w,l}=\left\lceil\frac{F}{W_l}\right\rceil.
\end{equation}
Since each \gls{prb} patch contributes $2|\mathcal{S}_p|$ tokens, a high-resolution window holds
\vspace{-4pt} 
\begin{equation}
    T_h=N_h\cdot 2|\mathcal{S}_p| = 24
\end{equation}
tokens, and a low-resolution window, after pooling by $\rho$, holds
\vspace{-7pt} 
\begin{equation}
    T_l=\frac{N_l}{\rho}\cdot 2|\mathcal{S}_p| = 24
\vspace{-7pt} 
\end{equation}
tokens.
Folding the windows of each sample into the batch dimension yields the branch inputs
\vspace{-4pt} 
\begin{equation}
    \mathbf{U}_h \in \mathbb{R}^{BN_{w,h}\times T_h\times P},
    \qquad
    \mathbf{U}_l \in \mathbb{R}^{BN_{w,l}\times T_l\times P}.
    \label{eq:low_res_tokens}
\vspace{-7pt} 
\vspace*{0.08in}
\end{equation}
Both branches therefore process a fixed sequence length, independent of $F$. When $F$ is not an exact multiple of a window size, the final window simply contains fewer valid tokens, and \gls{chea} processes only the available tokens in that window.







\vspace{-7pt} 

\subsection{High-Resolution Encoder}
\label{subsec:hr_encoder}
\vspace{-3pt} 

Each token of $\mathbf{U}_h$ is linearly embedded into a $d$-dimensional space and augmented with a positional encoding,
\vspace{-4pt} 
\begin{equation}
    \mathbf{Z}_h^{(0)}
    =
    \mathbf{U}_h\mathbf{W}_h^{\mathrm{in}}
    +
    \mathbf{b}_h^{\mathrm{in}}
    +
    \mathbf{E}_h
    \;\in\;
    \mathbb{R}^{BN_{w,h}\times T_h\times d},
\vspace{-4pt} 
\end{equation}
where $\mathbf{W}_h^{\mathrm{in}}\in\mathbb{R}^{P\times d}$ and $\mathbf{b}_h^{\mathrm{in}}\in\mathbb{R}^{d}$ are the patch-embedding weight and bias.
The learnable positional encoding $\mathbf{E}_h\in\mathbb{R}^{1\times T_h\times d}$ is shared by all windows and broadcast over the window--batch dimension.

The embedded tokens pass through a single pre-normalization transformer block, consisting of a self-attention sublayer followed by a feed-forward sublayer, each with a residual connection,
\vspace{-6pt} 
\begin{align}
  \hat{\mathbf{Z}}_h
    &=
    \mathbf{Z}_h^{(0)}
    +
    \mathrm{MHA}
    \!\left(
    \mathrm{LN}\!\left(\mathbf{Z}_h^{(0)}\right)
    \right),
    \label{eq:hr_encoder_attn}
    \\
    \mathbf{Z}_h
    &=
    \hat{\mathbf{Z}}_h
    +
    \mathrm{FFN}
    \!\left(
    \mathrm{LN}\!\left(\hat{\mathbf{Z}}_h\right)
    \right),
    \label{eq:hr_encoder_ffn}
\vspace{-9pt} 
\end{align}
where $\mathrm{LN}(\cdot)$ denotes layer normalization and $\hat{\mathbf{Z}}_h$ is the intermediate output after the attention sublayer.

The two sublayers act on a generic token sequence $\mathbf{Z}$, which in \eqref{eq:hr_encoder_attn}--\eqref{eq:hr_encoder_ffn} is the layer-normalized input of each sublayer.
Multi-head attention uses $N_{\mathrm{head}}$ heads with per-head dimension $d_a=d/N_{\mathrm{head}}$. For the $a$-th head,
\vspace{-4pt} 
\begin{equation}
    \mathbf{Q}_a=\mathbf{Z}\mathbf{W}^{Q}_a,\quad
    \mathbf{K}_a=\mathbf{Z}\mathbf{W}^{K}_a,\quad
    \mathbf{V}_a=\mathbf{Z}\mathbf{W}^{V}_a,
    \label{eq:qkv_proj}
\vspace{-4pt} 
\end{equation}
with $\mathbf{W}^{Q}_a,\mathbf{W}^{K}_a,\mathbf{W}^{V}_a\in\mathbb{R}^{d\times d_a}$.
The attention-weight matrix and the head output are
\vspace{-4pt} 
\begin{equation}
    \mathbf{A}_a
    =
    \mathrm{softmax}
    \!\left(
    \frac{\mathbf{Q}_a\mathbf{K}_a^{\mathrm{T}}}{\sqrt{d_a}}
    \right),
    \qquad
    \mathbf{O}_a
    =
    \mathbf{A}_a\mathbf{V}_a ,
\vspace{-4pt} 
\end{equation}
where $\mathbf{A}_a\in\mathbb{R}^{BN_{w,h}\times T_h\times T_h}$ has a fixed size $T_h\times T_h$, independent of $F$, and $\mathbf{O}_a\in\mathbb{R}^{BN_{w,h}\times T_h\times d_a}$.
The heads are concatenated and projected back to dimension $d$,
\vspace{-4pt} 
\begin{equation}
    \mathrm{MHA}(\mathbf{Z})
    =
    \mathrm{Concat}
    \!\left(
    \mathbf{O}_1,\ldots,\mathbf{O}_{N_{\mathrm{head}}}
    \right)
    \mathbf{W}^{O}
    \;\in\;
    \mathbb{R}^{BN_{w,h}\times T_h\times d},
\vspace{-4pt} 
\end{equation}
with $\mathbf{W}^{O}\in\mathbb{R}^{d\times d}$.
The feed-forward network is applied independently to each token,
\begin{equation}
    \mathrm{FFN}(\mathbf{Z})
    =
    \phi\!\left(\mathbf{Z}\mathbf{W}_1+\mathbf{b}_1\right)
    \mathbf{W}_2+\mathbf{b}_2,
\end{equation}
where $\mathbf{W}_1\in\mathbb{R}^{d\times d_{\mathrm{ff}}}$, $\mathbf{W}_2\in\mathbb{R}^{d_{\mathrm{ff}}\times d}$, $d_{\mathrm{ff}}$ is the hidden dimension, and $\phi(\cdot)$ is the GELU activation.
Both sublayers preserve the token-sequence shape $BN_{w,h}\times T_h\times d$.

Finally, the per-window token sequences are merged back along the frequency axis to form the high-resolution memory,
\vspace{-4pt} 
\begin{equation}
    \mathbf{M}_h
    =
    \mathrm{reshape}\!\left(\mathbf{Z}_h\right)
    \in
    \mathbb{R}^{B\times N_{w,h}T_h\times d}.
\vspace{-4pt} 
\end{equation}

\subsection{Low-Resolution Encoder}
\label{subsec:lr_encoder}

The low-resolution branch operates on the pooled windowed tokens $\mathbf{U}_l$.
The tokens are projected to the same embedding dimension $d$,
\vspace{-5pt} 
\begin{equation}
    \mathbf{Z}_l^{(0)}
    =
    \mathbf{U}_l\mathbf{W}_l^{\mathrm{in}}
    +
    \mathbf{b}_l^{\mathrm{in}}
    +
    \mathbf{E}_l
    \;\in\;
    \mathbb{R}^{BN_{w,l}\times T_l\times d},
\end{equation}
where $\mathbf{W}_l^{\mathrm{in}}\in\mathbb{R}^{P\times d}$ and $\mathbf{b}_l^{\mathrm{in}}\in\mathbb{R}^{d}$ are the patch-embedding weight and bias, and $\mathbf{E}_l\in\mathbb{R}^{1\times T_l\times d}$ is a learnable positional encoding.
The low-resolution encoder uses the same pre-normalization transformer structure as the high-resolution encoder, with its own independent weights,
\vspace{-2pt} 
\begin{align}
    \hat{\mathbf{Z}}_l
    &=
    \mathbf{Z}_l^{(0)}
    +
    \mathrm{MHA}
    \!\left(
    \mathrm{LN}\!\left(\mathbf{Z}_l^{(0)}\right)
    \right),
    \\
    \mathbf{Z}_l
    &=
    \hat{\mathbf{Z}}_l
    +
    \mathrm{FFN}
    \!\left(
    \mathrm{LN}\!\left(\hat{\mathbf{Z}}_l\right)
    \right).
\end{align}
The low-resolution memory is then formed by reshaping $\mathbf{Z}_l$,
\begin{equation}
    \mathbf{M}_l\in\mathbb{R}^{B\times N_{w,l}T_l\times d}.
\end{equation}

\subsection{Cross-Attention Decoder}
\label{subsec:decoder}

The decoder transfers the wideband context captured by the low-resolution branch back to the high-resolution pilot tokens through \emph{local} cross-attention.
Rather than letting every high-resolution token attend to every low-resolution token, \gls{chea} restricts attention to fixed-size groups.
Each low-resolution patch summarizes $\rho$ neighboring \glspl{prb}, so the decoder pairs that patch with the $\rho$ corresponding high-resolution \gls{prb} patches to form one group.
The number of groups equals the number of low-resolution patches, $N_g=\lceil F/(\rho P)\rceil=\lceil N_{\mathrm{PRB}}/\rho\rceil$.

Within a group, the high-resolution patches provide the query tokens and the single low-resolution patch provides the key/value tokens.
Since each \gls{prb} patch contributes $2|\mathcal{S}_p|$ tokens, the query and key/value lengths per group are
\begin{equation}
    T_q = \rho \cdot 2|\mathcal{S}_p| = 16,
    \qquad
    T_k = 2|\mathcal{S}_p| = 4 .
\end{equation}
Collecting the grouped tokens from the high- and low-resolution memories $\mathbf{M}_h$ and $\mathbf{M}_l$ yields
\begin{equation}
    \mathbf{G}_h \in \mathbb{R}^{BN_g \times T_q \times d},
    \qquad
    \mathbf{G}_l \in \mathbb{R}^{BN_g \times T_k \times d}.
\end{equation}

The grouped high-resolution tokens are mapped by an input projection and augmented with a learnable positional encoding,
\begin{equation}
    \tilde{\mathbf{G}}_h
    =
    \mathbf{G}_h\mathbf{W}_q^{\mathrm{in}}
    +
    \mathbf{b}_q^{\mathrm{in}}
    +
    \mathbf{E}_q ,
\end{equation}
where $\mathbf{W}_q^{\mathrm{in}}\in\mathbb{R}^{d\times d}$ and $\mathbf{b}_q^{\mathrm{in}}\in\mathbb{R}^{d}$ are the query projection weight and bias, and $\mathbf{E}_q\in\mathbb{R}^{1\times T_q\times d}$ is a positional encoding shared across all groups and broadcast over the group--batch dimension.
The decoder applies a single pre-normalization block with three sublayers, each wrapped in a residual connection: self-attention over the projected high-resolution tokens, cross-attention from these tokens to the low-resolution tokens, and a feed-forward network,

\begin{align}
    \mathbf{D}^{(1)}
    &=
    \tilde{\mathbf{G}}_h
    +
    \mathrm{MHA}_{\mathrm{self}}
    \!\left(
    \mathrm{LN}(\tilde{\mathbf{G}}_h)
    \right),
    \\
    \mathbf{D}^{(2)}
    &=
    \mathbf{D}^{(1)}
    +
    \mathrm{MHA}_{\mathrm{cross}}
    \!\left(
    \mathrm{LN}(\mathbf{D}^{(1)}),\,
    \mathbf{G}_l
    \right),
    \\
    \mathbf{D}^{(3)}
    &=
    \mathbf{D}^{(2)}
    +
    \mathrm{FFN}
    \!\left(
       \mathrm{LN}(\mathbf{D}^{(2)})
    \right),
\vspace{-5pt} 
\end{align}
where $\mathbf{D}^{(1)}$ and $\mathbf{D}^{(2)}$ are the intermediate outputs after the self-attention and cross-attention sublayers, and $\mathbf{D}^{(3)}$ is the block output.
The self-attention $\mathrm{MHA}_{\mathrm{self}}(\cdot)$ and the feed-forward network $\mathrm{FFN}(\cdot)$ follow the same definitions as in the high-resolution encoder (Section~\ref{subsec:hr_encoder}), with their own parameters.

In the cross-attention sublayer $\mathrm{MHA}_{\mathrm{cross}}(\cdot)$, for the $a$-th head, with queries from $\mathrm{LN}(\mathbf{D}^{(1)})$ and keys/values from $\mathbf{G}_l$,
\vspace{-3pt} 
\begin{equation}
    \mathbf{Q}_a = \mathrm{LN}(\mathbf{D}^{(1)})\mathbf{W}^{Q}_a,\quad
    \mathbf{K}_a = \mathbf{G}_l\mathbf{W}^{K}_a,\quad
    \mathbf{V}_a = \mathbf{G}_l\mathbf{W}^{V}_a,
\vspace{-4pt} 
\end{equation}
with $\mathbf{W}^{Q}_a,\mathbf{W}^{K}_a,\mathbf{W}^{V}_a\in\mathbb{R}^{d\times d_a}$ and $d_a=d/N_{\mathrm{head}}$.
The cross-attention-weight matrix and the head output are
\vspace{-5pt} 
\begin{equation}
    \mathbf{A}_a
    =
    \mathrm{softmax}
    \!\left(
    \frac{\mathbf{Q}_a\mathbf{K}_a^{\mathrm{T}}}{\sqrt{d_a}}
    \right),
    \qquad
    \mathbf{O}_a
    =
    \mathbf{A}_a\mathbf{V}_a ,
\vspace{-5pt} 
\end{equation}
where $\mathbf{A}_a\in\mathbb{R}^{BN_g\times T_q\times T_k}$ has a fixed size $T_q\times T_k=16\times 4$, and $\mathbf{O}_a\in\mathbb{R}^{BN_g\times T_q\times d_a}$.
The heads are concatenated and projected back to dimension $d$, giving $\mathrm{MHA}_{\mathrm{cross}}\in\mathbb{R}^{BN_g\times T_q\times d}$.

The decoder output $\mathbf{D}^{(3)}$ and the memory $\mathbf{M}_h$ are each projected to the patch dimension $P$ by $\mathbf{W}_{\mathrm{out}},\mathbf{W}_{\mathrm{skip}}\in\mathbb{R}^{d\times P}$ and reshaped to the common high-resolution patch layout, then
\vspace{-4pt} 
\begin{equation}
    \tilde{\mathbf{U}}_h
    =
    \mathbf{D}^{(3)}\mathbf{W}_{\mathrm{out}}
    +
    \mathbf{M}_h\mathbf{W}_{\mathrm{skip}}
    \;\in\;
    \mathbb{R}^{B\times \frac{F}{P}\times 2|\mathcal{S}_p|\times P}.
\vspace{-4pt} 
\end{equation}
The refined tokens $\tilde{\mathbf{U}}_h$ are then de-tokenized (reshaped) into the pilot-domain channel tensor
\vspace{-4pt} 
\begin{equation}
    \tilde{\mathbf{X}}_{p}
    \in
    \mathbb{R}^{F\times |\mathcal{S}_p|\times 2}.
\vspace{-3pt} 
\end{equation}

\vspace{-6pt} 

\subsection{Per-PRB Upsampling}
\label{subsec:upsampling}
\vspace{-4pt} 

The refined pilot tensor $\tilde{\mathbf{X}}_{p}\in\mathbb{R}^{F\times|\mathcal{S}_p|\times 2}$ is defined only on the $|\mathcal{S}_p|$ pilot symbols.
A single shared linear layer reconstructs the full slot \gls{prb} by \gls{prb}.
Slicing $\tilde{\mathbf{X}}_{p}$ along frequency gives, for each \gls{prb} $r$, a feature vector $\mathbf{u}_{r}\in\mathbb{R}^{48}$ collecting its $P\times|\mathcal{S}_p|\times 2$ entries, which is mapped to all $S$ symbols by
\vspace{-4pt} 
\begin{equation}
    \hat{\mathbf{u}}_{r}
    =\mathbf{W}_{\mathrm{up}}\mathbf{u}_{r}+\mathbf{b}_{\mathrm{up}}
    \in\mathbb{R}^{336},
    \qquad
    \mathbf{W}_{\mathrm{up}}\in\mathbb{R}^{336\times 48},
    \label{eq:prb_upsampling}
\vspace{-3pt} 
\end{equation}
where $336=P\times S\times 2$ for $P=12$ and $S=14$.
Reshaping each $\hat{\mathbf{u}}_{r}$ to $\mathbb{R}^{P\times S\times 2}$ and stacking the $F/P$ \glspl{prb} along frequency yields the full-slot estimate of link $\ell$,
\vspace{-3pt} 
\begin{equation}
    \hat{\mathbf{H}}_{\ell}\in\mathbb{R}^{F\times S\times 2},
    \label{eq:full_estimate}
\vspace{-3pt} 
\end{equation}
which estimates the target $\mathbf{H}_{\ell}^{\mathbb{R}}$ of~\eqref{eq:real_channel_tensor}.

\vspace{-3pt} 

\subsection{Stacked CHEA Stages}
\label{subsec:chea_stack}

\vspace{-1pt} 

The default \gls{chea} architecture stacks two pilot-domain refinement stages.
Let $\mathbf{X}^{(i)}_{p}$ denote the pilot-domain tensor at the input of the $i$th stage.
Each stage applies its transformation and updates the pilot-domain representation 
\vspace{-6pt} 
\begin{equation}
\mathbf{X}^{(i+1)}_{p}
=
\mathbf{X}^{(i)}_{p}
+
\alpha_i
\left(
\mathcal{F}_{\theta_i}\!\left(\mathbf{X}^{(i)}_{p}\right)
-
\mathbf{X}^{(i)}_{p}
\right),
\vspace{-6pt} 
\end{equation}
where $\mathcal{F}_{\theta_i}(\cdot)$ is the transformation implemented by the $i$th \gls{chea} stage and $\alpha_i$ is a trainable scalar that weights its residual correction.
The output of the final stage is passed to the per-\gls{prb} upsampling layer in \eqref{eq:prb_upsampling}, which reconstructs the full-slot channel estimate.




\subsection{Training CHEA}
\label{subsec:chea_training}
    
\vspace{-4pt} 

The model is trained to match the estimate $\hat{\mathbf{H}}_{\ell}$ of~\eqref{eq:full_estimate} to the target $\mathbf{H}_{\ell}^{\mathbb{R}}$ of~\eqref{eq:real_channel_tensor} by the Huber loss
\vspace{-8pt} 
\begin{equation}
    \mathcal{L}_{\mathrm{H}}
    =
    \frac{1}{2|\mathcal{D}|N_R N_{TX}}
    \sum_{\ell}\sum_{(f,s)\in\mathcal{D}}\sum_{c=1}^{2}
    h_{\delta}\!\big([\hat{\mathbf{H}}_{\ell}-\mathbf{H}_{\ell}^{\mathbb{R}}]_{f,s,c}\big),
    \vspace{-6pt} 
    \label{eq:huber_loss}
\end{equation}
where $c$ indexes the real/imaginary parts and $h_{\delta}(e)=\tfrac{1}{2}e^{2}$ for $|e|\le\delta$ and $\delta(|e|-\tfrac{1}{2}\delta)$ otherwise.
\vspace{-4pt} 

\subsection{Complexity and Scalability}
\label{subsec:chea_complexity}
\vspace{-1pt} 

\gls{chea} consists of \glspl{matmul}, GELU activation, and bias operations. 
For the default two-stage \gls{chea} stack with $d=16$, $d_{\mathrm{ff}}=2\times d$, $N_h=6$, $N_l=24$, $\rho=4$, and $|\mathcal{S}_p|=2$, \gls{chea} requires $6496F$ \glspl{mac} with approximately $36$K trainable parameters. 
Larger variants can be obtained by increasing $d$, or $d_{\mathrm{ff}}$. We make \gls{chea}-XL by setting $d=64$, and keeping other parameters the same as \gls{chea}.

The computational complexity per link and parameter count of the neural estimators evaluated in this paper are summarized in Table~\ref{tab:complexity_comparison}. 
Attention-based methods, HA02, Channelformer, and CEViT include terms that scale quadratically with the number of subcarriers $F$, mainly due to global attention operations or bandwidth-dependent dense mappings. 
In contrast, all CHEA variants scale linearly with $F$ because the attention operations are confined to fixed-size local windows and the final reconstruction is performed per \gls{prb}. 
The convolution-based model, InterpolationNet, also has linear complexity. However, its MAC coefficient is considerably larger than that of CHEA. 
In particular, InterpolationNet requires approximately $9.5\times$ more \glspl{mac} than \gls{chea} for any value of $F$.

\begin{table}[t]

\centering
\caption{Comparison of MACs and parameters.}
\label{tab:complexity_comparison}
\vspace{-6pt} 

\begin{tabular}{lcc}
\hline
\textbf{Models} & \textbf{MACs} & \textbf{Params} \\
\hline
HA02 & $68F^{2} + 312F$ & $20F^{2} + 28F + 55$ \\
Channelformer & $344F^{2} + 52{,}296F$ & $18F^{2} + 28F + 22381$ \\
CEViT & $28F^{2} + 68{,}700F$ & $45F + 203730$ \\
InterpolationNet & $61{,}776 F$ & $56F + 5,410 $ \\
CHEA & $6{,}496F$ & $36\,\mathrm{K}$ \\
CHEA-XL & $65{,}984F$ & $274\,\mathrm{K}$ \\
\hline
\end{tabular}
\vspace{-17pt} 
\end{table}

\vspace{-5pt} 

\section{Simulation Results}
\label{sec:simulation_results}
\vspace{-4pt} 

All models are implemented in Python/TensorFlow, and evaluated on two bandwidths: $10$ PRBs ($F=120$ subcarriers) and $22$ PRBs ($F=264$ subcarriers). 

\vspace{-5pt} 

\subsection{Simulation Setup}
\label{subsec:simulation_setup}
\vspace{-4pt} 

The end-to-end uplink transmission is simulated with a standard-compliant 5G NR PUSCH chain using Sionna~\cite{hoydis2022sionna}.  
The transmitter side consists of two single-layer uplink streams.  Each stream is mapped to one dual-polarized antenna element, resulting in two transmitted layers.  
The base station is equipped with two dual-polarized antenna elements, corresponding to four receive antenna elements.  
This gives the $4\times2$ MIMO configuration used throughout the evaluation. 
The carrier frequency is set to $2$ GHz and the modulation and coding configuration follows MCS index 14, i.e., 16-QAM.

Training samples are generated from the \gls{3gpp} \gls{umi} channel model. Each training sample corresponds to an independent random drop, which randomizes the user positions, propagation geometry, angles of arrival and departure, path delays, and path powers. The user speed is sampled in the interval $[0,34]$ m/s.

The test set is generated from the \gls{3gpp} \gls{tdl}-A channel model. The delay spread is randomly selected between $10$ ns and $300$ ns, and the Doppler shift is uniformly chosen between $0$ Hz and $325$ Hz.


\vspace{-4pt} 

\subsection{Training Procedure}
\label{subsec:training_procedure}
\vspace{-4pt} 

All neural channel estimators are trained with the Adam optimizer and the same staged learning-rate schedule on an \gls{snr} range of $[0, 25]$dB.
The first training phase uses $50$K iterations with learning rate $10^{-3}$, the second phase uses $100$K iterations with learning rate $10^{-4}$, and the final phase uses $200$K iterations with learning rate $10^{-5}$. 
The Huber loss defined in~\eqref{eq:huber_loss} is used with $\delta=1$ for training.
For InterpolateNet, HA02, Channelformer, and CEViT, separate models are trained for each evaluated bandwidth.
Hence, each of these baselines has one model for the 10-PRB case and another model for the 22-PRB case.  
In contrast, CHEA is trained once with a random number of allocated PRBs drawn from $1$ to $24$.

\vspace{-4pt} 

\subsection{MSE Performance}
\label{subsec:results_mse_bandwidths}
\vspace{-2pt} 

\begin{figure}[t]
\vspace{0.03in}
\centerline{\includegraphics[width=.95\columnwidth]{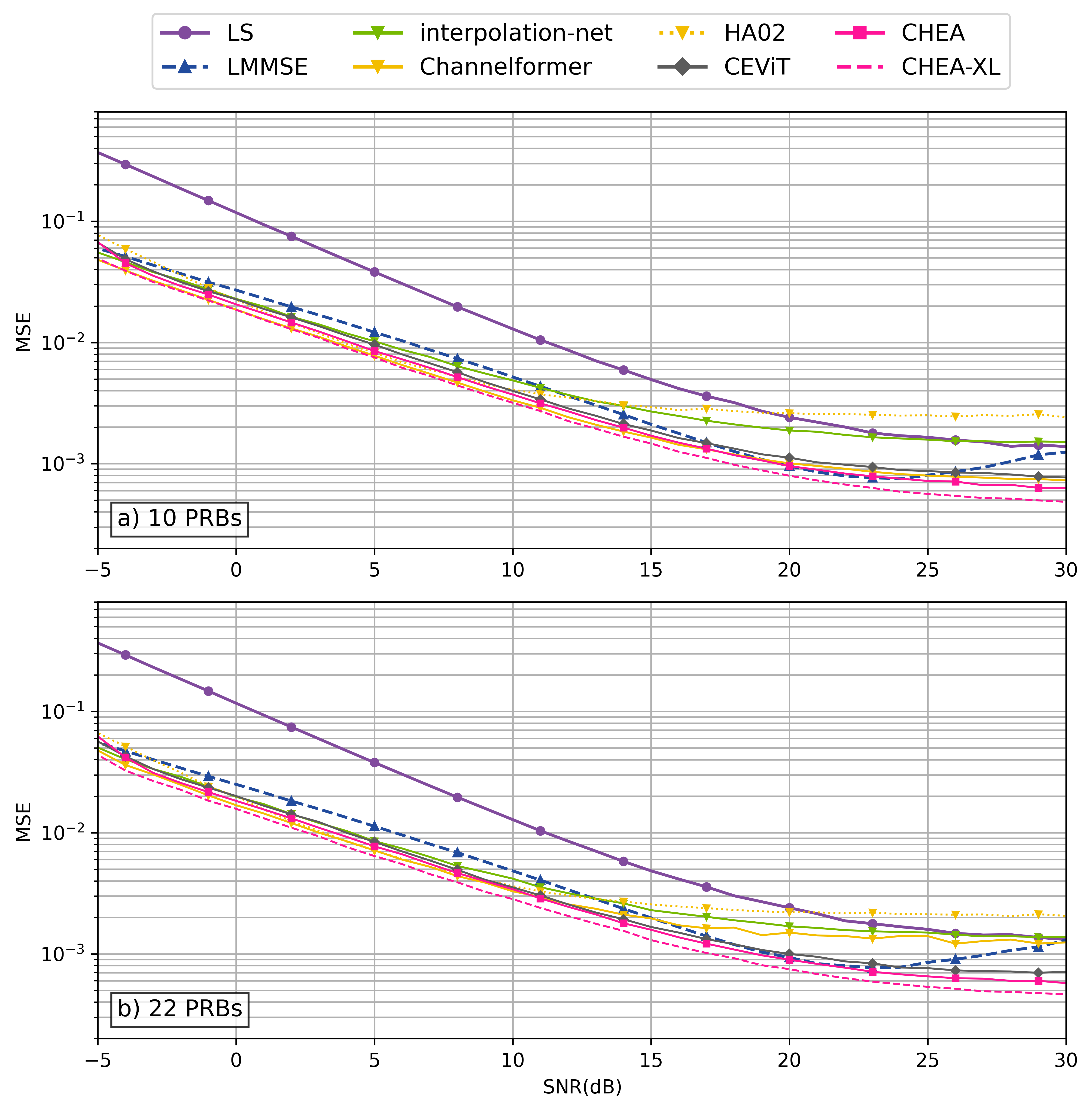}}
\vspace{-13pt} 
\caption{MSE performance with a)10 and b)22 PRBs.}
\label{fig:img_tdla_mse}
\vspace{-15pt} 
\end{figure}

Fig.~\ref{fig:img_tdla_mse}a, and~\ref{fig:img_tdla_mse}b show the MSE performance on the \gls{tdl}-A test channel for the 10-PRB and 22-PRB bandwidths, respectively. 
The same general trends are observed in both cases. 
The LS estimator has the highest MSE over the full SNR range because it relies only on pilot-domain division and interpolation. 
The LMMSE estimator substantially improves over LS by exploiting covariance information. 
However, its performance saturates at high SNR and exhibits a mild upturn.



The neural baselines reduce the error floor compared with LS and LMMSE, but their high-SNR performance depends strongly on the architecture. 
InterpolationNet and Channelformer improve the low-to-medium SNR region, while HA02 shows a visible high-SNR floor. 
CEViT achieves a lower high-SNR MSE than these baselines.
In both bandwidths, CHEA and CHEA-XL provide the best overall MSE performance.
The CHEA model already achieves the lowest or near-lowest MSE over most of the SNR range, while CHEA-XL further reduces the high-SNR error floor.

The proposed architecture remains effective as the bandwidth increases. 
CHEA improves channel-estimation accuracy while also simplifying deployment across different \gls{pusch} allocations. 
Together with the complexity results in Table~\ref{tab:complexity_comparison}, these results demonstrate that CHEA provides a favorable accuracy--complexity trade-off: it achieves strong MSE performance while preserving linear scaling with the number of subcarriers.

\vspace{2pt} 

\section{Conclusion}
\label{sec:conclusion}

This paper proposed CHEA, a scalable attention-based channel estimator for 5G NR MIMO PUSCH. 
CHEA replaces global attention over the full resource grid with a multi-resolution windowed design, where a cross-attention decoder transfers the low-resolution context to high-resolution tokens.

By restricting attention to fixed-size windows and performing the final reconstruction on a per-\gls{prb} basis, the proposed architecture scales linearly with the number of subcarriers and supports different \gls{prb} allocations using a single trained model.
Simulation results on \gls{3gpp} \gls{tdl}-A channels show that \gls{chea} achieves lower \gls{mse} than conventional estimators and state-of-the-art neural channel estimation baselines, while maintaining significantly lower complexity.
These results demonstrate that multi-resolution local attention provides an effective, scalable, and flexible solution for practical neural channel estimation in \gls{5g} \gls{nr} uplink systems. 
Future work will include a detailed component-wise ablation study, an analysis of the learned attention patterns, and an end-to-end link-level evaluation.

\vspace{6pt} 


\Urlmuskip=0mu plus 1mu\relax
\def\UrlBreaks{\do\/\do-}
\bibliographystyle{IEEEtran_nodash}
\bibliography{IEEEabrv,bib}

\end{document}